\documentclass[proof]{WileyASNA-v1}

\articletype{Article Type}%

\begin{document}

\title{Fusion dynamics of $^{12}$C+$^{12}$C reaction: An astrophysical interest within the relativistic mean-field approach}

\author[1]{Shilpa Rana*}
\author[1]{Raj Kumar}
\author[2,3,4]{M. Bhuyan}

\authormark{Shilpa Rana, Raj Kumar, and M. Bhuyan}
\address[1]{School of Physics and Materials Science, Thapar Institute of Engineering and Technology, Patiala 147004, India}
\address[2]{Department of Physics, Faculty of Science, University of Malaya, Kuala Lumpur 50603, Malaysia}
\address[3]{Atomic Molecular and Optical Research Group, Advanced Institute of Materials Science, Ton Duc Thang University, Ho Chi Minh City, Vietnam}
\address[4]{Faculty of Applied Sciences, Ton Duc Thang University, Ho Chi Minh City, Vietnam}

\corres{*Shilpa Rana, \email{shilparana1404@gmail.com}}


\abstract{The $^{12}$C+$^{12}$C fusion reaction plays a significant role in the later phases of stellar evolution. For a better understanding of the evolution involved, one must understand the corresponding fusion-fission dynamics and reaction characteristics. In the present analysis, we have studied the fusion cross-section along with the S-factor for this reaction using the well-known M3Y and recently developed R3Y nucleon-nucleon (NN) potential along with the relativistic mean-field densities in double folding approach. The density distributions and the microscopic R3Y NN potential are calculated using the NL3$^*$ parameter set. The $\ell$- summed Wong model is employed to investigate the fusion cross-section, with $\ell_{max}$-values from the sharp cut-off model. The calculated results are also then compared with the experimental data. It is found that the R3Y interaction gives a reasonable agreement with the data.
}


\jnlcitation{\cname{\author{Shilpa Rana}, \author{Raj Kumar}, and  \author{M. Bhuyan}} (\cyear{2020}), \ctitle{Fusion dynamics of $^{12}$C+$^{12}$C reaction: An astrophysical interest within the relativistic mean-field approach}, \cjournal{Astronomische Nachrichten}, \cvol{XXXX; YY:xxxx}.}


\maketitle

\footnotetext{\textbf{Abbreviations:} RMF, Relativistic mean field; NN, Nucleon-Nucleon}

\section{Introduction}
\label{sec1}
Nuclear fusion reactions provide the source for stellar energy. In the process of stellar Helium burning, the main products are $^{12}$C and $^{16}$O. For massive stars, with mass greater than $8M_\odot$, the Carbon and Oxygen burning reactions, and principally the $^{12}$C+$^{12}$C, plays a significant role in later phases of stellar evolution and explosions (\citet{patterson1969,aguilera2006,tang20}). The crucial temperature for such reactions to occur in the stars lies in the range of 0.8 to 1.2 GK, which corresponds to the center-of-mass energies ($E_{c.m.}$) of 1-3 MeV. The experimental measurement of the fusion cross-section at such low energies of astrophysical interest is tedious due to the suppression of cross-section by the Coulomb barrier (\citet{ass2013,patterson1969}). In addition to this, there are resonant structures observed even at a very low energy region. As a result, large uncertainties persist in reaction rate while extrapolating the data at an astrophysically significant energy range (\citet{tang20,beck20}). So the experimental measurement of the fusion cross-section for $^{12}$C+$^{12}$C fusion reaction have been limited to the energies above $E_{c.m.}=2.1$ MeV (\citet{beck20,zhang20}). 

To extrapolate the data in the lower energy regions of astrophysical interest, the theoretical modeling of heavy-ion $^{12}$C+$^{12}$C fusion reaction is necessary. Various phenomenological and microscopic models have been developed to explain the fusion dynamics of these heavy-ion reactions (\citet{beck20,zhang20}). Further, to remove the effects arising due to the Coulomb barrier, the fusion cross-section for astrophysical reactions is defined in terms of astrophysical S-factor. This S-factor contains all the intrinsic nuclear factors which influence the reaction cross-section and is observed to follow a rising trend towards the lower energies (\citet{patterson1969,aguilera2006,zhang20}). The barrier penetration model using the proximity adiabatic $\&$ Krappe-Nix-Sierk potentials (\citet{aguilera2006}), coupled channel calculations (\citet{esb2011,ass2013}),  density-constrained time-dependent Hartree-Fock method (\citet{umar2012}) and time-dependent wave-packet method (\citet{torres2018}) have been used to study the fusion cross and S-factor of Carbon and Oxygen burning fusion reactions. In this direction, it will be interesting and also important to determine the applicability of the relativistic mean-field model with the most popular M3Y and recently developed R3Y NN-interaction potential for the study of $^{12}$C + $^{12}$C reaction system. 

This paper is organized as follows: In Sec. \ref{sec2}, we discuss the theoretical formalism used for the calculations of total interaction potential, fusion cross-section, and astrophysical S-factor. Sec. \ref{sec3} is assigned to the discussion of the results. Finally, a summary and a brief conclusion are given in Sec. \ref{sec4}.

\section{Theoretical Formalism}
\label{sec2}
The total interaction potential between two interacting nuclei is given by 
\begin{eqnarray} 
V_T(R)=V_C(R)+V_n(R)+V_\ell(R).
\label{vt}
\end{eqnarray}
Here, $V_C (R) = Z_p Z_t e^2/R$ and $V_\ell(R) = \frac{\hbar^2\ell(\ell+1)}{2\mu R^2}$ represent the well known repulsive Coulomb and centrifugal potentials respectively. The term $V_n (R)$ is the short-range attractive nuclear potential. The resultant of Coulomb and nuclear potentials gives the fusion barrier. We have adopted the double folding approach (\citet{satc79}) to obtain the nuclear interaction potential $V_n (R)$ and is given as,
\begin{eqnarray}
\vspace{-1.5cm}
V_{n}(\vec{R}) = \int\rho_{p}(\vec{r}_p)\rho_{t}(\vec{r}_t)V_{eff}
\left( |\vec{r}_p-\vec{r}_t +\vec{R}| {\equiv}r \right) 
d^{3}r_pd^{3}r_t. 
\label{vnn}
\end{eqnarray}
Here, $\rho_p$ and $\rho_t$ are the nuclear density distributions of interacting projectile and target nuclei respectively. $V_{eff}$ is the effective nucleon nucleon (NN) interaction potential. 

The densities of the projectile and target nuclei in Eq. (\ref{vnn}) are obtained using relativistic mean field (RMF) formalism. More details of RMF approach can be found in the Refs. (\citet{ring96,bhuy18,bhuyan2020,lala09} and references therein). Here we have used two kinds of nucleon-nucleon potentials, namely, (1) the most popular M3Y potential given in terms of three Yukawa terms (\citet{satc79,bhuy18,bhuyan2020}); and (2) the recently developed relativistic R3Y potential by solving the RMF equations of motion for mesons in limit of one-meson exchange (\citet{sing12,bhuy18,bhuyan2020,bidhu14}). For the present study we have used recently developed non-linear NL3$^*$ parameter set (\citet{lala09}), which is the refitted version of the NL3 parameter set. It is worth mentioning that the relativistic R3Y NN potential is analogous to the M3Y potential and can be used for various nuclear studies, such as proton and cluster radioactivity, nuclear decay, nuclear fusion and so on (\citet{sing12,bhuy18,bhuyan2020,bidhu14}). The barrier characteristics i.e. barrier height, position and frequency are extracted from the total interaction potential and are used to estimate the fusion cross-section for $^{12}$C+$^{12}$C system. 

 Wong gave a formula that use s-wave barrier characteristics (\citet{wong73}) to obtain the cross-section for fusion. It excludes the angular momentum dependence of potential, which was later included by (\citet{gupta09}). The extended formula is named as $\ell$-summed Wong model (\citet{gupta09,bhuy18,bhuyan2020}). In this model, the fusion cross-section in terms of the partial wave is given as, 
\begin{eqnarray}
\sigma(E_{c.m.})=\frac{\pi}{k^{2}} \sum_{\ell=0}^{\ell_{max}}(2\ell+1)P_\ell(E_{c.m}).
\label{crs}
\end{eqnarray}
Here, $E_{c.m}$ is the center-of-mass energy of two colliding nuclei and $P_{\ell}$ is known as the transmission coefficient for $\ell^{th}$ partial wave. It is generated using Hill-Wheeler approximation (\citet{wong73} and references therein). In terms of barrier height $V_{B}^{\ell}(E_{c.m.})$ and curvature $\hbar\omega_{\ell}(E_{c.m.})$, $P_{\ell}$ is written as,
\begin{eqnarray}
P_{\ell}=\Bigg[1+exp\bigg(\frac{2 \pi V_{B}^{\ell}(E_{c.m.})-E_{c.m.}}{\hbar \omega_{\ell}(E_{c.m.})}\bigg)\bigg]. 
\end{eqnarray}
$P_{\ell}$ describes the penetration of barrier given by Eq. (\ref{vt}). The $\ell_{max}$-values are obtained from the sharp cut-off model (\citet{beck81}) and extrapolated for below barrier energies. It is to be noted that this model can be used to calculate the fusion cross-section around the Coulomb barrier. At energies far below the barrier, the Coulomb force dominates. So to remove most of the Coulomb barrier penetration effect, the astrophysical S-factor was introduced. It depends upon the intrinsic effects of nuclear forces and for $^{12}$C+$^{12}$C system it is given by (\citet{patterson1969,aguilera2006}),
\begin{eqnarray}
S = \sigma  E_{c.m.}  exp(87.21 E^{-1/2}+0.46E).
\label{sf}
\end{eqnarray}
\section{Results and Discussions}
\label{sec3}
The main aim of the present work is to test the relativistic mean-field approach in terms of density and NN-interaction potential for the fusion cross-section of $^{12}$C+$^{12}$C, which holds a great astrophysical significance, and then use the same approach to predict the cross-section at below barrier energies of astrophysical interest as no experimental data is available. 
\begin{figure}
\centering
\vspace{-0.5cm}
\includegraphics[scale=.35]{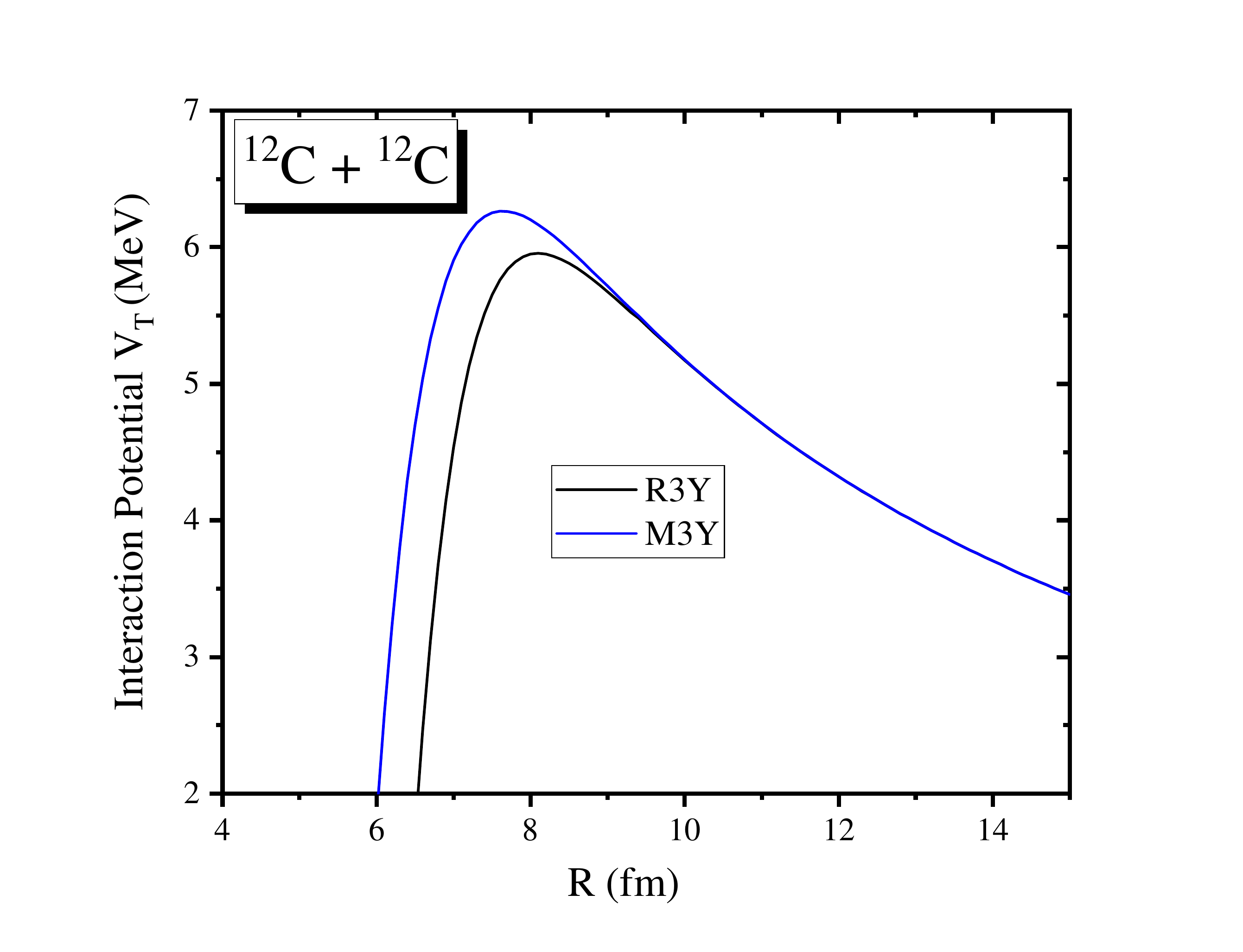}
\caption{Total interaction potential ($V_T(R)=V_N(R) +V_C(R)$), at $\ell$=0, as a function of radial separation R for $^{12}$C+$^{12}$C system using M3Y (blue) and R3Y (black) NN interaction.}
\label{fig1}
\end{figure}
\begin{figure}
\centering
\vspace{-0.5cm}
\includegraphics[scale=.35]{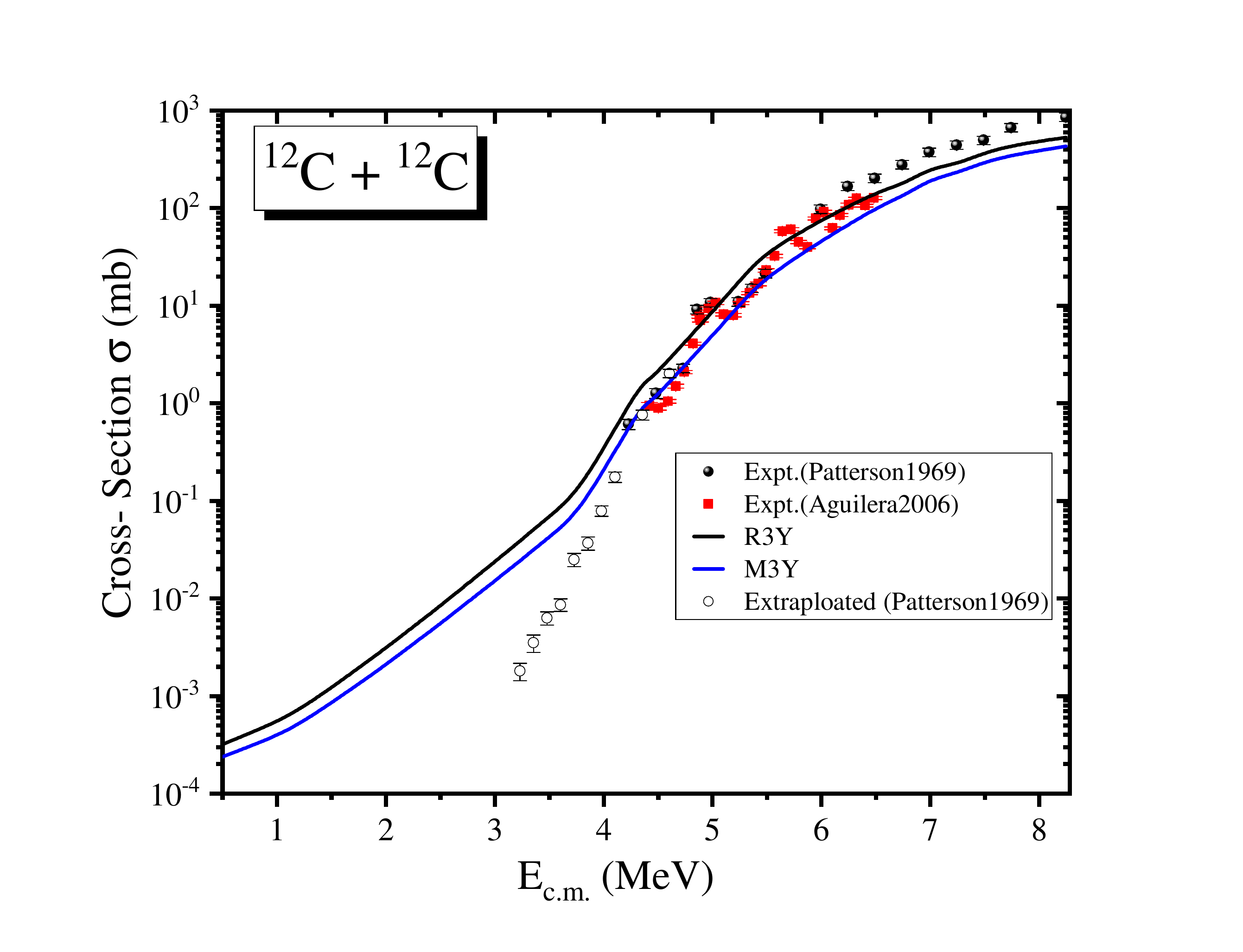}
\caption{Fusion cross section $\sigma$ (mb) as function of center of mass energy $E_{c.m.}$ (MeV) for $^{12}$C+$^{12}$C system using R3Y (black) and M3Y (blue) NN interaction. The experimental and extrapolated data are taken from Refs. (\citet{patterson1969,aguilera2006}).}
\label{fig2}
\end{figure}
\begin{figure}
\centering
\vspace{-0.5cm}
\includegraphics[scale=.35]{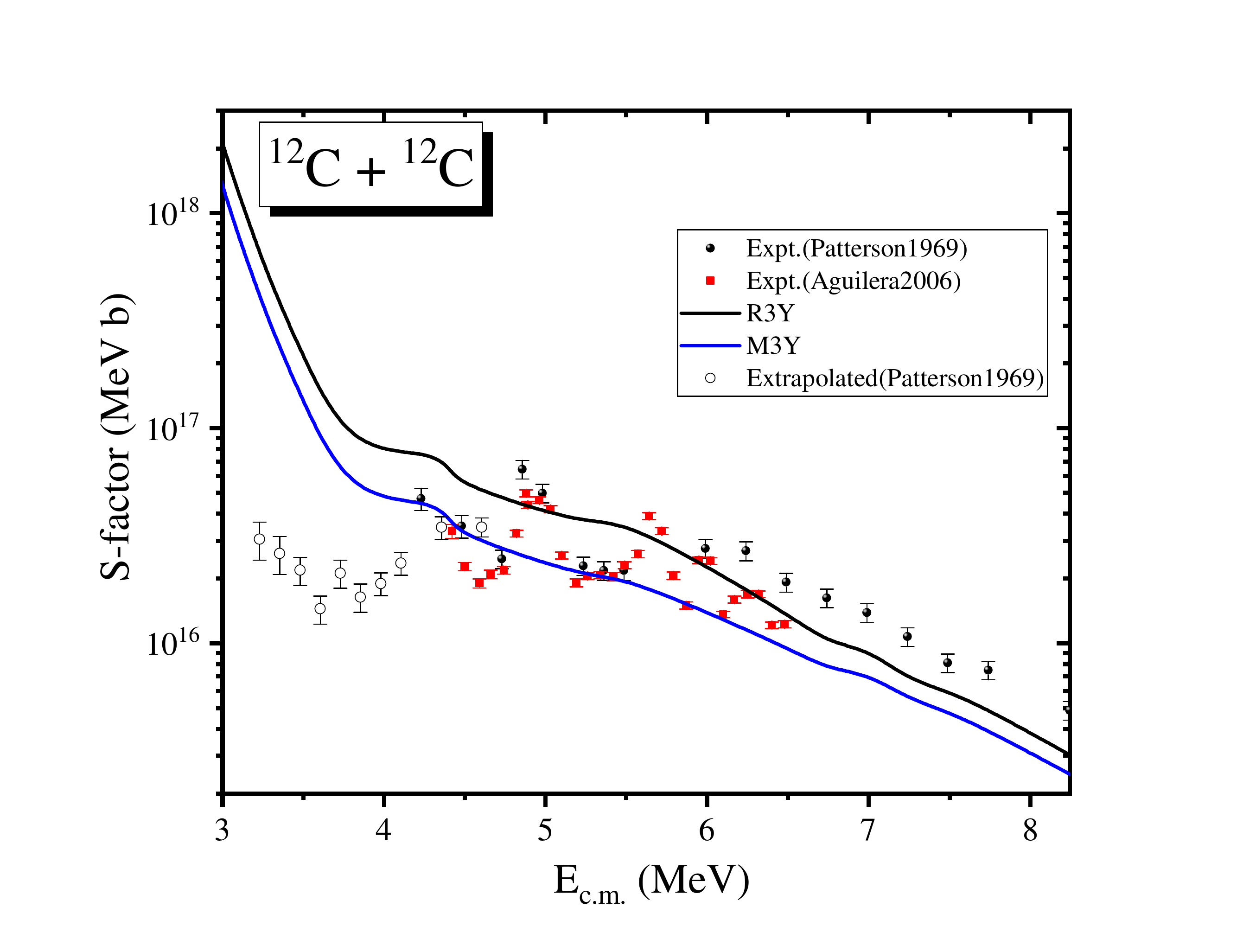}
\caption{Same as Fig. \ref{fig2} but for the astrophysical S-factor.}
\label{fig3}
\end{figure}

To study the fusion dynamics of the astrophysical system we have obtained RMF densities and nucleon-nucleon potential for NL3$^*$ parameter set. The densities for the interacting nuclei are then folded with the phenomenological M3Y and relativistic R3Y NN- potential to estimate the nuclear interaction potential. Fig. \ref{fig1} shows the barrier region of total interaction potential as a function of radial separation R for $^{12}$C + $^{12}$C system. The blue and black lines in the figure are for nuclear interaction potential obtained by using M3Y and R3Y  NN potentials, respectively. It is observed from Fig. \ref{fig1} that M3Y NN interaction gives comparatively higher fusion barrier height as compared to  R3Y potential. The values of barrier height $V_B$ for M3Y and R3Y NN interactions are 6.264 MeV and 5.955 MeV respectively. The barrier position $R_B$ for R3Y is shifted a bit towards the right (more radial distance) as compared to the M3Y one. This shows that the R3Y interaction is more attractive than the M3Y interaction,  which will be reflected in the fusion characteristics.

\noindent
The fusion cross-section as a function of center-of-mass energy ($E_{c.m.}$) is presented in Fig. \ref{fig2} for both M3Y (blue line) and R3Y (black line) NN interactions. The calculations of the fusion cross-section are started from $E_{c.m.}$=0.5 MeV onward. This energy range is far below the observed Coulomb barrier and lies well within the astrophysically significant energy range (1-3 MeV). The $\ell_{max}$ values at above barrier regions are calculated from the sharp cut off model (\cite{beck81}) and extrapolated for below barrier energies. For comparison the experimental (black spheres) and the extrapolated (black circles) data from Ref. (\citet{patterson1969}) and experimental data (solid red squares) from Ref. (\citet{aguilera2006}) is also plotted in the Fig. \ref{fig2}. It can be observed here that R3Y NN interaction gives a higher cross-section as compared with the M3Y interaction potential at all energy regions. On comparing the calculated results with the experimental data it is observed that R3Y NN interaction gives a better fit to the experimental data than the M3Y interaction at below barrier as well as above barrier region. If we observe the far below barrier region ($E_{c.m.}< 4.0$ MeV) then it is found that both M3Y and R3Y NN interaction potentials give higher fusion cross-section as compared to the extrapolated ones (\citet{patterson1969}), as no experimental data is available for this energy range. It is to be noted here that the $\ell_{max}$ can take only integer values else a relatively better fit to the data could also be achieved with R3Y interaction. So overall, the calculated fusion cross-section with R3Y NN interaction gives a nice fit to the measured experimental data and overestimates the extrapolated data of (\citet{patterson1969}). 

\noindent
The astrophysical S-factor calculated using Eq. (\ref{sf}) for R3Y (black line) and M3Y (blue line) NN-interactions is compared with the experimental (solid black dots from Ref. (\citet{patterson1969}) and red squares from Ref. (\citet{aguilera2006}) and extrapolated data (hollow circles from Ref. (\citet{patterson1969}) as a function of $E_{c.m.}$ for $^{12}$C+$^{12}$C reaction in Fig. \ref{fig3}. It is observed that the astrophysical S-factor increases sharply at far below barrier energy. As evident from the fusion cross-section as well, the S-factor corresponding to R3Y NN interaction potential follows a similar trend of the experimental data. The calculated S-factor at far below barrier energies is higher for both the cases of M3Y and R3Y potentials as compared to the extrapolated fusion cross-section (\citet{patterson1969}). This shows that either R3Y and M3Y potentials give relatively attractive nuclear interaction potential and/or modification of Eq. (\ref{sf}) by implementing proper structural input. 
\vspace{-0.5 cm}
\section{Summary and Conclusions}
\label{sec4}
We have calculated the fusion cross-section and S-factor for $^{12}$C+$^{12}$C system which holds a great astrophysical significance. The nuclear interaction potential is obtained from the double folding model furnished with relativistic mean-field (RMF) density distributions along with phenomenological M3Y and microscopic R3Y NN interaction. The $\ell$-summed Wong formula is employed to calculate the fusion cross-section. The fusion cross-section and S-factor obtained using microscopic R3Y NN interaction derived from RMF theory is found to be more close to the experimental data as compared to phenomenological M3Y NN interaction both at around the barrier as well as above energy regions. At far below barrier center of mass energies ($E_{c.m.}< 4.0$ MeV), both M3Y and R3Y NN interaction potentials are observed to give higher values of fusion cross-section as well the S-factor as compared to the available extrapolated ones. Thus the present work has great motive to test the relativistic mean-field approach for giving a reasonable fit to the fusion cross-section of $^{12}$C+$^{12}$C, 
and then use the same approach to predict the cross-section at below barrier energies of astrophysical interest as no experimental data is available. 
It will be of future interest to investigate the fusion of other Carbon and Oxygen burning reactions such as $^{12}$C+$^{16}$O and $^{16}$O+$^{16}$O etc. within this microscopic approach, which can lead to a better understanding of fusion cross-section at lower energies of astrophysical interest and hence the later phases of stellar evolution. 
\vspace{-0.5 cm} 
\section*{Acknowledgments}
This work was supported by \fundingAgency{DAE-BRNS Project Sanction No. 58/14/12/2019-BRNS, FOSTECT Project Code: FOSTECT.2019B.04, and FAPESP Project Nos. 2017/05660-0}.

\bibliography{Wiley-ASNA}%



\end{document}